\documentclass[prb,aps,letter,twocolumn,amsmath,amssymb,superscriptaddress,floatfix]{revtex4-2}

\usepackage[utf8]{inputenc}
\usepackage{setspace}
\usepackage{graphicx}
\usepackage{color}
\usepackage{soul}
\usepackage{hyperref}
\usepackage{miller}
\usepackage{pdfpages}
\usepackage{bbold}
\usepackage{gensymb}
\usepackage{physics2}
\usephysicsmodule{ab,ab.braket}
\usepackage{siunitx}
\usepackage{orcidlink}

\newcommand{\simlt}
      {\ifmmode       { \raisebox{-.8em}{$<$}\atop\sim}
         \else        {$\raisebox{-.8em}{$<$}\atop\sim$}
      \fi}

\makeatletter
\def\@hangfrom@section#1#2#3{\@hangfrom{#1#2}#3}
\def\@hangfroms@section#1#2{#1#2}
\makeatother

\makeatletter
\AtBeginDocument{\let\LS@rot\@undefined}
\makeatother

\usepackage[version=4]{mhchem}

\begin{document}


\title{How to measure loop currents in scanning tunneling microscopy}

\author{Victoire Morisseau}
\affiliation{SUPA, School of Physics and Astronomy, University of St Andrews, North Haugh, St Andrews, KY16 9SS, United Kingdom}
\author{Luke C. Rhodes \orcidlink{0000-0003-2468-4059}}
\affiliation{SUPA, School of Physics and Astronomy, University of St Andrews, North Haugh, St Andrews, KY16 9SS, United Kingdom}
\author{Peter Wahl \orcidlink{0000-0002-8635-1519}}
\affiliation{SUPA, School of Physics and Astronomy, University of St Andrews, North Haugh, St Andrews, KY16 9SS, United Kingdom}
\affiliation{Physikalisches Institut, Universit\"{a}t Bonn, Nussallee 12, 53115 Bonn, Germany}
\author{Carolina A. Marques \orcidlink{0000-0002-3804-096X}}
\affiliation{SUPA, School of Physics and Astronomy, University of St Andrews, North Haugh, St Andrews, KY16 9SS, United Kingdom}

\date{\today}

\begin{abstract}
The emergence of loop current phases, where spontaneous loops of orbital currents give rise to a weak local magnetic moments, has been proposed to exist in a number of quantum materials based on measurements that pick up weak signatures of time reversal symmetry breaking or small magnetic moment order. The most prominent example is as an explanation of the pseudogap phase on the underdoped side of the phase diagram of the high-temperature cuprate superconductors, but more recently, it has been proposed to occur in Kagome materials and at the surface layer of \ce{Sr2RuO4}. Experimental results have, however, been inconclusive so far, some detecting signatures that can be understood as emerging due to loop current phases, whilst others have not detected any significant proof. One of the techniques that should be able to pick up local signatures of loop current orders is low temperature scanning tunneling microscopy and spectroscopy (STM/STS), however firm predictions of how to detect them are missing. Here, we provide specific predictions for how loop current orders in a square lattice can be seen in spectroscopic maps, using models of the cuprate high-temperature superconductors and of the surface layer of \ce{Sr2RuO4}. We find that, besides lifting degeneracies at the specific ordering vector of the loop current order, a finite spin polarisation emerges when spin-orbit coupling is present, signatures of which can be detected in spin-polarised STM.

\end{abstract}


\maketitle

The pseudogap phase of the cuprate superconductors remains a major unresolved mystery that may hold the key to an understanding of high-temperature superconductivity. There is a wide range of proposals for its theoretical description, ranging from a precursor phase with pairs but no macroscopic phase coherence\cite{emery_importance_1995,randeria_precursor_1997} to proposals for competing phases\cite{timusk_pseudogap_1999}. A competing phase that could account for some of the phenomena seen in the pseudogap phase, notably the existence of small local moments, is the staggered loop current phase\cite{chakravarty_hidden_2001,varma_theory_2006}. Since the original proposal of this exotic state for the description of the pseudogap phase and reports of its detection in the cuprates\cite{kaminski_spontaneous_2002,mangin-thro_intra-unit-cell_2015}, it has been suggested that this phase is stabilized in a number of materials, including the Kagome material family\cite{mielke_time-reversal_2022,guo_switchable_2022}, iridates\cite{jeong_time-reversal_2017} and in the surface of \ce{Sr2RuO4}\cite{fittipaldi_unveiling_2021,mazzola_signatures_2024}. Despite these promising leads, unequivocal demonstration of the occurrence of the loop current state has been difficult, and its existence in some of these materials has been disputed\cite{croft_no_2017,li_no_2022}. Signatures of the loop current phase include the detection of local moments in Nuclear magnetic resonance (NMR), neutron diffraction or small moment magnetism in  muon spin spectroscopy ($\mu$SR). In angle-resolved photoemission spectroscopy (ARPES), the loop current order is expected to result in an asymmetry of the dichroic signal\cite{varma_proposal_2000}, where the related signals are however often small. For scanning tunneling microscopy (STM), which can probe the density of states and its spatial variation with high energy resolution, one of the difficulties in searching for evidence of loop current orders has been a scarcity of clear predictions of what their signatures in tunneling spectra or spatial maps of the local density of states would be. Previous work has typically concentrated on the signatures in cuprate superconductors and the interplay of the pseudogap phase, superconductivity and a possible loop current order.\cite{varma_proposal_2000,allais_loop_2012}

Here, we perform realistic simulations of how loop currents would be detected in an STM experiment. Using the continuum Green's function method\cite{choubey2014,Kreisel2015,kreisel_towards_2016,kreisel_quasi-particle_2021,wahl_calcqpi_2025} (see section~\ref{supp:Methods} for details), we simulate quasiparticle interference (QPI) measurements by calculating the local density of states (LDOS) as a function of position and energy for models with and without loop current orders. We start by identifying the effects of introducing loop currents for a one band tight-binding model of the cuprate superconductors\cite{Kreisel2015}, and then move on to introduce them for a multiband system with and without spin-orbit coupling, using a model of the surface layer of \ce{Sr2RuO4}, covering two materials where signatures for loop current orders have been reported\cite{kaminski_spontaneous_2002,fittipaldi_unveiling_2021,mazzola_signatures_2024}. We here neglect possible self energy effects that could accompany the formation of the loop current order.\cite{varma_proposal_2000} While the calculation of self energy corrections is beyond the scope of the present work, inclusion in the continuum calculation would be straightforward.\cite{rhodes_revealing_2026} Our work provides predictions of the signatures of loop currents in spectroscopy measurements by STM as well as by spin-polarized STM, and specific limits for the strength of the loop currents for their signal to be experimentally detectable.

\begin{figure}[!ht]
    \begin{center}
        \includegraphics[width=\columnwidth]{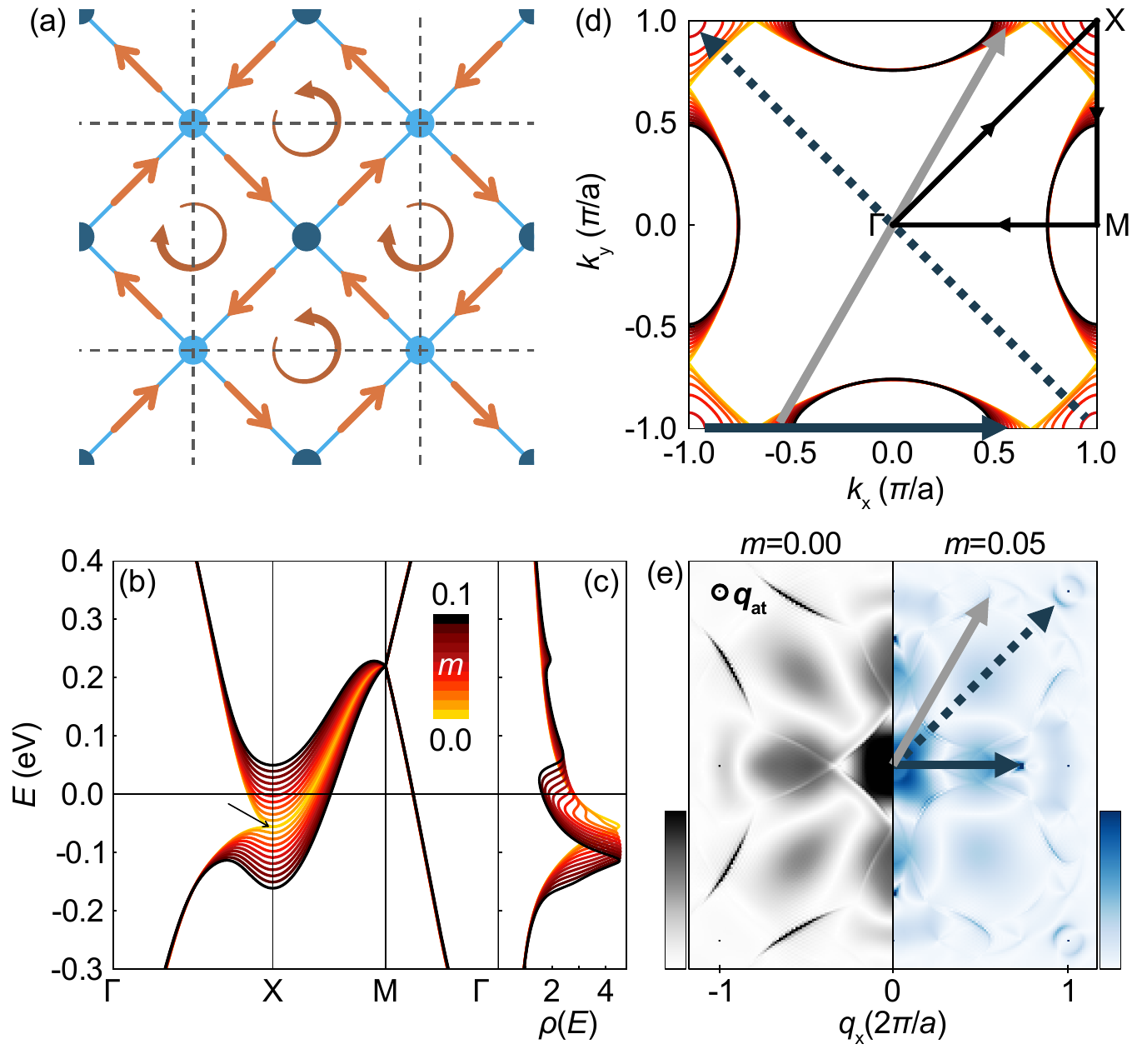}
    \end{center}
    \caption{\textbf{Loop currents in a cuprate model.} (a) Sketch of a tight-binding model on a square lattice with a two-atom unit cell and loop currents, representative of the cuprate model.\cite{Kreisel2015} The loop currents enclose the one-atom unit cell and form a staggered order (orange arrows). The dark and light blue circles indicate the position of the two distinct \ce{Cu} atoms in the unit cell. (b) Electronic band structure along the path $\Gamma$-$X$-$M$-$\Gamma$ and (c) total density of states $\rho(E)$ obtained for the model in (a) for loop currents corresponding to increasing magnetic moment, from $m=0.0$ (yellow) to $0.1$ (black). The black arrow in (b) points to the VHs that splits at the $X$-point with increasing loop current strength. (d) Fermi surface of the same models, showing the closing of the Fermi pockets at the $X$-point with increasing $m$. The black lines indicate the path along which (b) was taken. (e) Constant energy QPI maps $\tilde{g}(\mathbf{q},V)$ for loop currents of $m=0.0$ (grey) and $m=0.05$ (blue) at $E=\qty{0}{\eV}$. The circle indicates the atomic peaks $q_{\mathrm{at}}$. The arrows indicate characteristic scattering vectors not present without loop currents. The same arrows are shown in (d).}
    \label{fig:fig1}
\end{figure}

Multiple geometries of loop currents have been proposed for the cuprate and ruthenate materials, often with intra-unit-cell loop currents running between the transition metal (Cu or Ru) and the oxygen atoms surrounding it.\cite{varma1997non, varma_theory_2006} Here, for simplicity, we consider loop currents covering the whole unit cell, with the current orbiting in the opposite direction in the adjacent unit cells as introduced in ref.~\cite{chakravarty_hidden_2001} and depicted in Fig.~\ref{fig:fig1}(a), without any assumption about their origin or stability. This staggered loop current order results in a doubling of the unit cell, but no net orbital moment.  The generalisation to other geometries of the
loop currents is straightforward.

Starting from a tight-binding Hamiltonian $H_0$, the loop currents are added to the nearest neighbour hopping terms along a specific bond through $\hat{H}_\mathrm{LC}$,\cite{zotos1997transport} 
\begin{equation}
\label{eq:hamilton}
   \hat{H} = \hat{H}_0+\hat{H}_\mathrm{LC},
\end{equation}
where $\hat{H}_\mathrm{LC}=i\sum_\mathrm{NN}(-1)^{n+m}t_\mathrm{LC}(\hat{c}_{0,0}^\dagger \hat{c}_{n,m} + h.c.)$ and $\hat{c}_{n,m}^{\dagger}$ and $\hat{c}_{n,m}$ are the creation and annihilation operators, respectively, and the sum is over the nearest neighbours (NN).
The current $I$ between two neighbouring sites $n$ and $n+1$ is given by
\begin{equation}
I = \frac{e}{\hbar} \times t_\mathrm{LC}.
\end{equation}
It is contained in the imaginary part of the hopping term $t=t_0e^{i\phi_{\mathrm{LC}}}$ in $\hat{H}$, where $t_0$ is the nearest neighbour hopping without loop currents, and $\phi_{LC}$ is the loop current phase defined such that $t_{\mathrm{LC}}=t_0 \sin\phi_{LC}$, guaranteeing that the magnitude of $|t|$ is preserved when the loop currents are included in the model.

To estimate what the magnitude of the imaginary part of the hopping term $t$ is for the signatures of loop currents reported from neutron diffraction and $\mu$SR, we consider the values of the magnetic dipole moments, $m$, obtained from neutron diffraction\cite{fauque_magnetic_2006,bourges_loop_2022} and $\mu$SR\cite{fittipaldi_unveiling_2021}. Typical values for the ordered local moment are $m\mu_\mathrm{B}\sim (0.01\ldots 0.1)\mu_\mathrm{B}$, with induced local fields of order $B\approx 10 \mathrm{G}$. Using the magnetic moment of a magnetic dipole, we  can calculate the value of the loop currents that would produce such magnetic dipoles through 

\begin{equation}
    m\mu_B  = I\cdot A, 
\end{equation}  

where $\mu_B$ is the Bohr magneton, $I$ is the current, and $A = 14.4$~\AA$^2$ is the approximate area of a closed loop of Cu or Ru atoms in a square lattice (bond lengths $\sim 3.8$~\AA). The magnetic moment is thus given by

\begin{equation}
    m=\frac{e A}{\mu_{\mathrm{B}}\hbar} t_0 \sin\phi_{\mathrm{LC}}.
\end{equation}

The smallest estimate using $m=0.01$ thus gives $t_\mathrm{LC} = 2.6\mathrm{m eV}$, and for the largest value of $m$ gives $t_\mathrm{LC} = 26\mathrm{m eV}$. 
This is consistent with a bond current of $6.4 \mathrm{\mu A}$ as previously estimated for the cuprates.\cite{chakravarty_hidden_2001}

For the calculation of QPI, we perform continuum LDOS calculations (see section~\ref{supp:Methods} for details), and introduce a point-like potential scatterer. We neglect changes to the loop current order due to the defect since those changes are expected to be localized\cite{zheng_quasiparticle_2025} and the QPI patterns are dominated by contributions far away from the scatterer.

We first discuss how loop currents change the electronic structure and QPI patterns for a tight-binding model of a high-$T_\mathrm{C}$ cuprate superconductor, using the hopping parameters and \ce{Cu}-$d_{x^2-y^2}$ Wannier function from ref.~\cite{Kreisel2015} (details in section~\ref{supp:Methods}). The tight-binding model is for a single orbital on a square lattice. The staggered loop currents as illustrated in Fig.~\ref{fig:fig1}(a) have an ordering vector of $\mathbf{q}=(1,1)$ (in units of $\pi/a$, where $a$ is the lattice constant of the one-atom unit cell), resulting in a doubling of the unit cell and in a hybridization gap at the $X$-point of the Brillouin zone that splits the Van Hove singularity (VHs) by $\Delta E_{\mathrm{VHs}}$ for increasing $m$, shown in Fig.~\ref{fig:fig1}(b). As a result, a partial energy gap $\Delta E$ appears in the density of states (DOS), Fig.~\ref{fig:fig1}(c). The magnitude of $\Delta E_{\mathrm{vHs}}$ is proportional to the strength of the loop currents, with $\Delta E_{\mathrm{VHs}}\sim m\cdot \qty{2.11}{\eV}$. For the typical magnitudes of the magnetic moment reported from experiments of about $m=0.01$, the corresponding partial energy gap in the DOS due to the loop currents would be on the order of \qty{20}{\milli\eV}, which would be easily detected by scanning tunneling spectroscopy at temperatures around $T=\qty{4.2}{\K}$.

The splitting of the Van Hove singularity with the loop currents results in clear changes to the Fermi surface, where the Fermi pockets around the $X$-point of the Brillouin zone decrease in size for increasing values of $m$, disappearing above the Fermi level for large enough values. These changes are reflected in the QPI patterns, as seen in Fig.~\ref{fig:fig1}(e), where we compare the QPI calculated for the models without and with loop currents. For the case with loop currents (right panel) additional scattering vectors are seen, indicated by blue and grey arrows in Fig.~\ref{fig:fig1}(e), in particular close to the atomic peaks and in the direction of the characteristic vector of the staggered order (dotted blue arrow).

\begin{figure}[!ht]
    \begin{center}
        \includegraphics[width=\columnwidth]{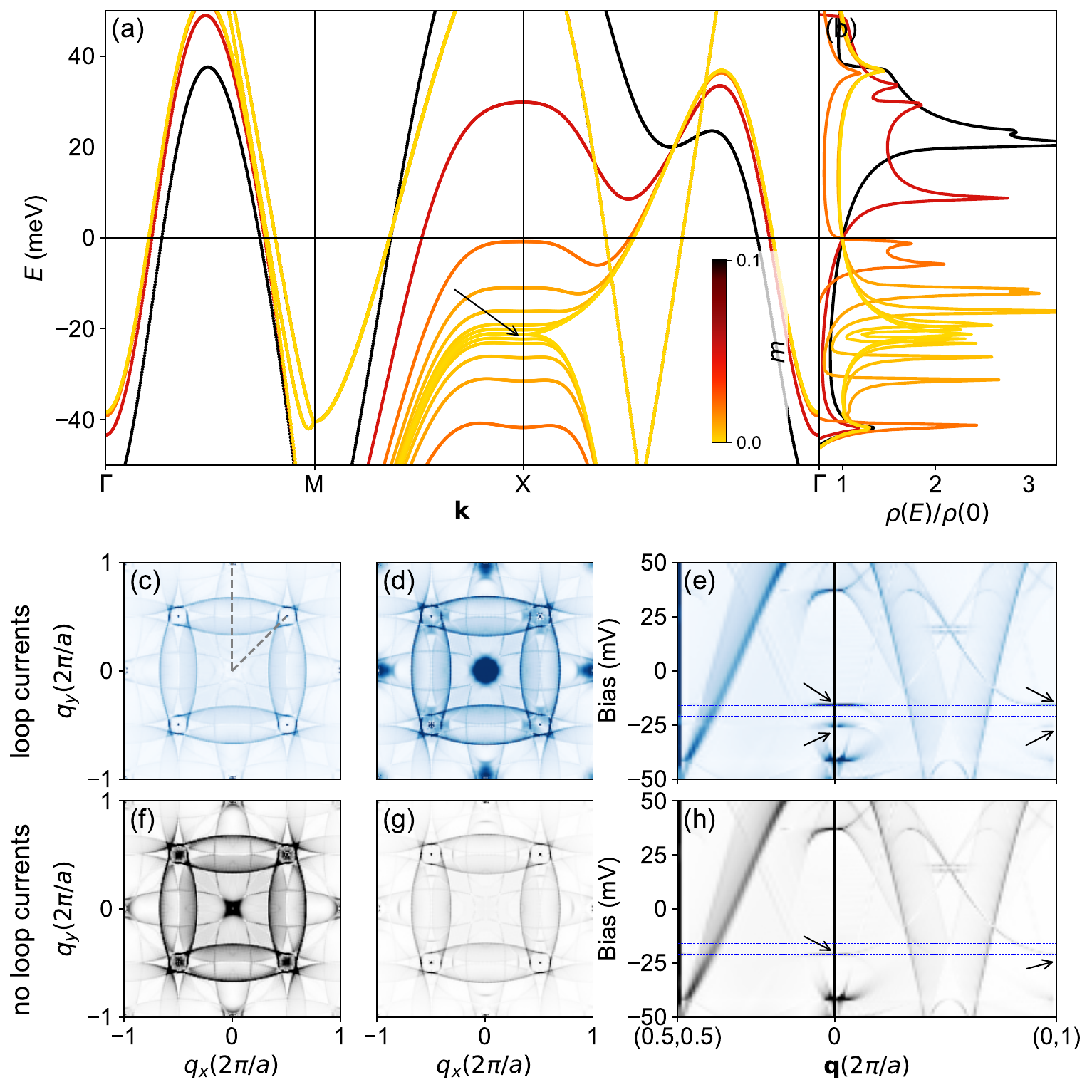}
    \end{center}
    \caption{\textbf{Loop currents in \ce{Sr2RuO4} and their signatures in QPI.}  (a) Electronic band structure of \ce{Sr2RuO4} in a 2-atom unit cell including loop current order. We show the band structure for loop currents with a strength of $m=0$, $0.001$, $0.002$, $0.005$, $0.01$, $0.02$, $0.05$, and $0.1$, respectively. The loop current order has been added only on the $d_{xy}$ band. A clear splitting of the $d_{xy}$ VHs can be seen which increases with increasing strength of the loop currents. (b) Normalized density of states $\rho(E)/\rho(0)$ corresponding to the same models, also showing the opening of a partial gap, with a width which increases similar to the splitting of the $d_{xy}$ VHs. (c, d) Constant energy maps $\tilde{g}(\mathbf{q},V)$ of the QPI for $V_1=-21\mathrm{mV}$ and $V_2=-16\mathrm{mV}$ for a QPI map with loop currents corresponding to an ordered moment of $m=0.005$. The energies were chosen at the energy of the $d_{xy}$ VHs without loop currents and its upper branch with loop currents (compare (b)). (d) shows a cut through the QPI along the dashed lines in (c). Panels (f-h) show the same calculation but with no loop currents.}
    \label{fig:fig2}
    \label{fig:fig3}
\end{figure}

To investigate how the loop currents affect a multi-orbital model with spin-orbit coupling, we follow the same procedure as for the one-band model and introduce the loop current order illustrated in Fig.~\ref{fig:fig1}(a) in a tight-binding model of the surface of \ce{Sr2RuO4}. By doing this, we aim to establish how the loop current order reported in the surface layer of \ce{Sr2RuO4}\cite{fittipaldi_unveiling_2021,mazzola_signatures_2024} can be detected using scanning tunneling microscopy and spectroscopy.

We start from a tight-binding model for the surface layer of \ce{Sr2RuO4} obtained from Density-functional theory (DFT), which already has two atoms per unit cell due to the octahedral rotations.\cite{chandrasekaran_engineering_2024} The unit cell hence remains the same when introducing the loop current order. To describe the electronic structure in the vicinity of the Fermi energy, it is sufficient to consider the $t_{2g}$ manifold. For the model calculations shown here, we added the loop currents only to the $d_{xy}$ orbital. Adding them to all orbitals does not modify the results significantly, because only the $d_{xy}$ orbital has a Van Hove singularity at the $X$-point in proximity to the Fermi energy that becomes affected by the loop currents.

Fig.~\ref{fig:fig2}(a) shows the resulting electronic structure. The impact of the loop currents on the electronic structure is most clearly seen at the zone corner, at the $X$-point, where the $d_{xy}$ Van Hove singularity splits proportionally to the strength of the loop currents. As in the cuprate model, this splitting results in a partial energy gap in the density of states, Figure~\ref{fig:fig2}(b). The $d_{xy}$ VHs is in the vicinity of the Fermi energy so that moderate values of $m$ are sufficient to split the VHs enough to push one VHs across the Fermi level, Figure \ref{fig:fig2} (b).

The formation of the partial gap due to the splitting of the VHs is probably the most tangible experimental signature that can be detected in an STM measurement, as it is expected to result in signatures in the differential conductance as well as in QPI patterns. 
To provide verifiable predictions for experimentally accessible quantities, we have simulated the impact that loop currents have on the QPI patterns in \ce{Sr2RuO4}. Figure \ref{fig:fig3}(c-h) show simulated QPI maps $\tilde{g}(\mathbf{q},eV)$ of the quasi-particle interference with (c-e) and without (f-h) the loop current order. The patterns are fairly similar to one another however show differences in intensity when comparing QPI energy layers at the same energy. Comparing panels in Fig.~\ref{fig:fig3}(f,g) with Fig.~\ref{fig:fig3}(c,d), respectively, clearly shows a high-intensity at $q=(0,0), (1,0), (0.5,0.5)$, characteristic of the scattering vectors from a VHs at the X-point, however, they appear at different energies, in Fig.~\ref{fig:fig3}(f) without loop current and in Fig.~\ref{fig:fig3}(d) with loop currents, consistent with the splitting of the VHs. The difference between the two cases is most clearly seen in the energy cut through the QPI, Fig.~\ref{fig:fig3}(e, h), where the splitting of the $d_{xy}$ VHs can be identified by the dispersions of the scattering vectors near $\mathbf{q}=(0,0)$ (highlighted by arrows).

\begin{figure}[!ht]
    \begin{center}
        \includegraphics[width=\columnwidth]{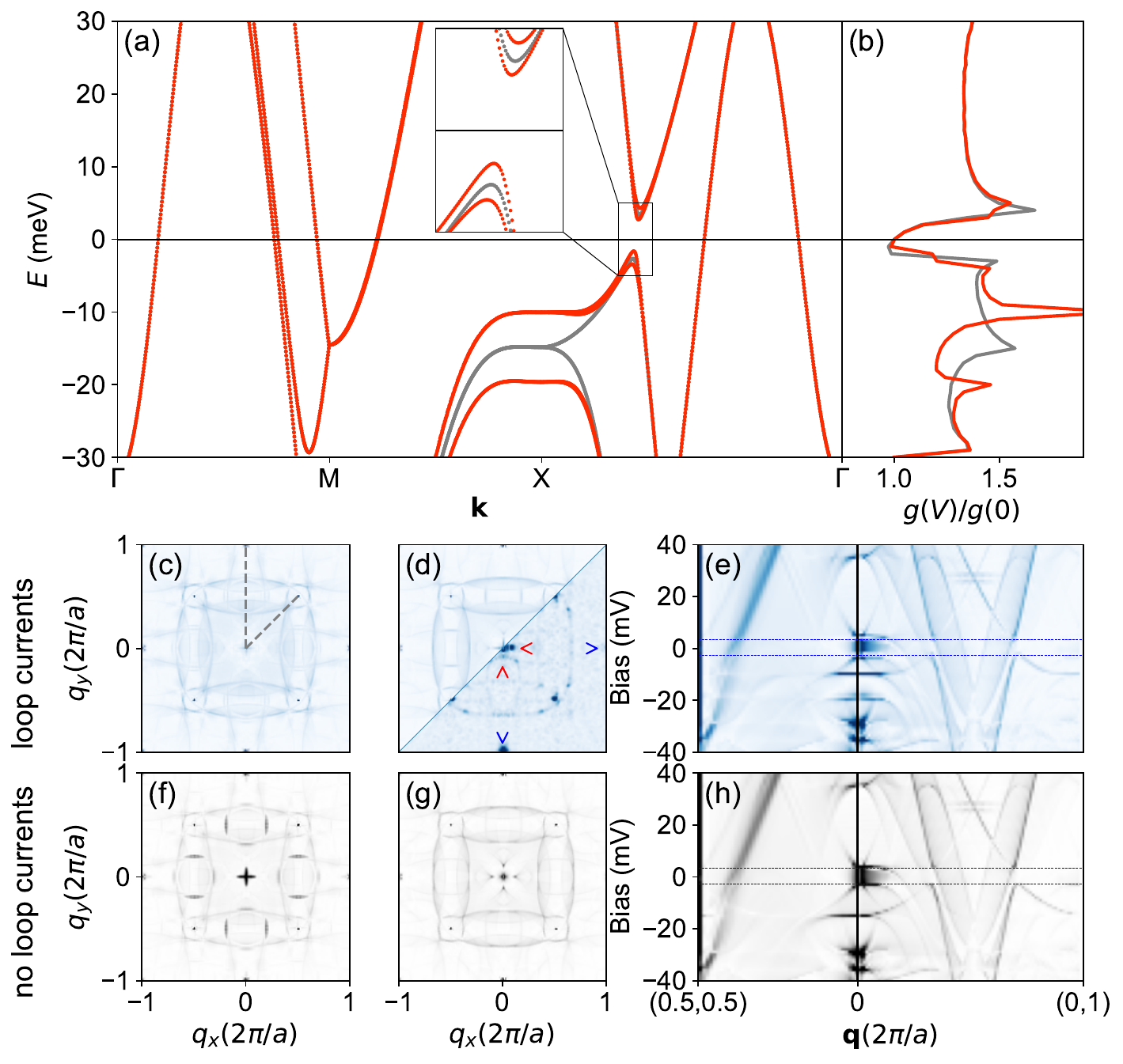}
    \end{center}
    \caption{\textbf{Signatures of loop currents in QPI including SOC.} (a, b) Band structure and cLDOS for a model of the surface of \ce{Sr2RuO4} including spin-orbit coupling and loop currents with $m=0.005$. The SOC-induced hybridization gap close to the Fermi energy can be clearly seen. The loop currents result in a small splitting of about $2\mathrm{meV}$ at the gap edges of the hybridization gap. (c, d) Constant energy map $\tilde{g}(\mathbf{q},V)$ of the QPI for $V_1=-2.8\mathrm{mV}$ and $V_2=3.36\mathrm{mV}$ for a QPI map with loop currents corresponding to an ordered moment of $m=0.005$. In (d), the lower right half shows an experimental QPI image at the same voltage from ref.~\cite{kreisel_quasi-particle_2021} for comparison. Blue and red arrowheads indicate symmetry-equivalent positions where the QPI differs, demonstrating nematicity of the electronic states. (e) energy cut through the QPI along the dashed lines in (c). Panels (f-h) show the same calculation but with no loop currents. }
    \label{fig:qpiwsoc}
\end{figure}

The electronic structure of \ce{Sr2RuO4} shows strong influence from spin-orbit coupling, which shows up as hybridization gaps close to the Fermi energy between bands of $d_{xy}$ and $d_{xz/yz}$ orbital character\cite{haverkort_strong_2008,chandrasekaran_engineering_2024}. The calculations so far have neglected the influence of the atomic spin-orbit coupling. In Fig.~\ref{fig:qpiwsoc} we show the band structure and a QPI calculation including atomic spin-orbit coupling of \qty{150}{meV} at the Ru site. The band structure now shows a hybridization gap near the Fermi energy, Fig.~\ref{fig:qpiwsoc}(a). The introduction of the loop current order results in a small splitting of the edges of the SOC-induced hybridization gap, inset in Fig.~\ref{fig:qpiwsoc}(a), seen also in the density of states as a kink in the gap near the Fermi energy, Fig.~\ref{fig:qpiwsoc}(b). This splitting is a consequence of the broken time-reversal and mirror symmetries introduced by the loop current order: The loop currents correspond to a non-local orbital moment and an associated angular momentum which couples, via the spin-orbit coupling, to the spin. The staggered loop current order hence results in a spin-splitting at the edges of the SOC-induced hybridization gap.

In panels Fig.~\ref{fig:qpiwsoc}(c-h), we show the corresponding quasi-particle interference for energies around the hybridization gap, where from experiment the QPI is strongest, including a comparison with experiment in Fig.~\ref{fig:qpiwsoc}(d). The differences are most readily seen in the energy line cut of the QPI, figs.~\ref{fig:qpiwsoc}(e, h), where the edges of the SOC induced hybridization gap are split when loop currents are included (Fig.~\ref{fig:qpiwsoc}(e)), while they are not split in the case without loop currents (Fig.~\ref{fig:qpiwsoc}(h)). This shows a clear experimentally detectable signature of the loop current order. 

\begin{figure}[!ht]
    \begin{center}
        \includegraphics[width=\columnwidth]{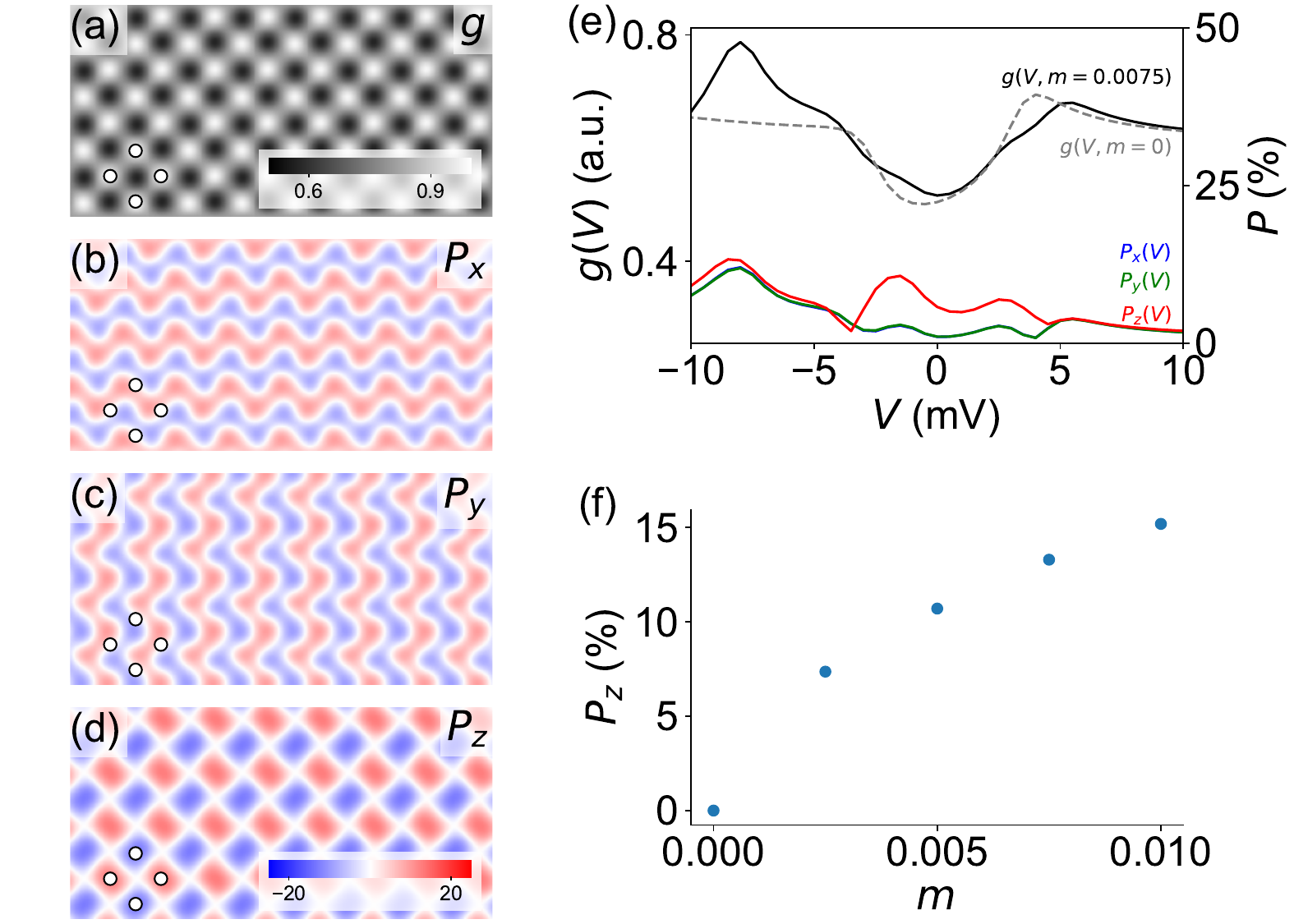}
    \end{center}
    \caption{\textbf{Detection of loop currents using spin-polarized STM.}  (a-d) Map of the continuum DOS $g(\mathbf{r},V)$ for \qty{8}{\milli\V} (a) and the spin polarization along the $x$ (b), $y$ (c) and $z$ (d) axes at the same energy. Clear spin polarization can be observed. Green circles indicate positions of Sr atoms, which are the atoms seen in atomically resolved topographic images. (e) Continuum density of states $g(V)$ and spin polarization $P(V)$ as a function of bias voltage $V$, showing strongest spin polarization at the position of the $d_{xy}$ Van Hove singularity and at the edges of the hybridization gap. (f) Maximal spin polarization $P_z$ as a function of strength of the ordered moment in the loop current phase. The energy broadening was set to \qty{1}{\milli\eV}, corresponding to the thermal broadening at \qty{4.2}~K.}
    \label{fig:fig4}
\end{figure}

The spin-splitting of the edges of the SOC gap due to the loop currents leads to a locally staggered spin-polarization that could in principle be detected in spin-polarised STM (SP-STM). Our continuum LDOS calculations allow us to quantify the expected spin-polarization and simulate the resulting spatial patterns. To estimate the magnitude of the loop currents that lead to a detectable spin-polarization $P$, we look at the spin-projection in real space. While the spin-averaged real space local density of states shows the atomic lattice, indistinguishable from the model without loop currents, the spin projections in real space onto the $x$, $y$ and $z$ axes show characteristic patterns due to the presence of loop currents, as seen in Fig.~\ref{fig:fig4}(b-c) for $m=0.0075$. The spin-projections onto the in-plane axes show a stripe pattern that changes sign between lines of adjacent atoms and runs along the respective $x$ and $y$ directions, Fig.~\ref{fig:fig4}(b, c). The spin-projection onto the $z$-axis has the dominant magnitude, showing a checkerboard order, with adjacent atoms having opposite signs. The spin-polarization $P$ is of the order of $\sim 15\%$, which could be detected by SP-STM. The overall spin-polarization $P$ changes as a function of energy, Fig.~\ref{fig:fig4}(e), where the largest difference between $P_z$ and $P_{x,y}$ is observed at the edges of the SOC gap, and the maximum magnitude of $P$ occurs at the $d_{xy}$ Van Hove singularity.
We investigated how the spin-polarization $P_z$ changes as a function of the staggered moment, Fig.~\ref{fig:fig4}(f). The spin-polarization stays below $\sim 15\%$, but the signal is large enough to be detected even for $m$ less than half of the reported moments. The dependence of the spin polarization on the loop currents moment is a consequence of the energy broadening of the electronic states, here set to $1$~meV corresponding to the expected thermal broadening at $4.2$~K, but it shows that as soon as the edges of the SOC-induced hybridization gap exhibit an appreciable splitting, a sizeable spin polarization can be detected.

Our work combines theoretical models of loop current order at the tight-binding level with realistic modelling of STM measurements using the continuum Green's function approach to predict measurable signatures of loop current orders. We show that for the reported magnitudes of the loop current order, clear signatures in tunneling spectra and QPI are expected. However, the specific signatures depend on both details of the underlying electronic structure as well as on the specific symmetry of the loop current order. This result underpins the need for microscopic calculations to compare with the experimental data.

For both the cuprates and the ruthenates, the staggered loop current order results in a splitting of a VHs at the Brillouin zone corner due to band folding induced by the loop current ordered moment $\textbf{q}=(1,1)$ (see Fig.~\ref{fig:fig1},~\ref{fig:qpiwsoc}). This splitting is directly proportional to the strength of the loop currents identified by the magnetic moment $m$. It results in a partial energy gap in the DOS of the order of 10-20meV for the reported strength of the loop currents in both systems, which should be clearly detectable in tunneling spectra, however does not represent a signature that is unique to a loop current order. The signatures in QPI are more subtle, with extra scattering vectors appearing at $\mathbf{q}=(0,0)$ and $(1,0)$ due to the splitting of the VHs. 

Introducing loop currents in a model of the surface of \ce{Sr2RuO4} allowed us to distinguish signatures of loop currents with and without inclusion of spin-orbit coupling. Without spin-orbit coupling, the main effect of the loop currents is the splitting of the $d_{xy}$ VHs, while accounting for SOC leads to a subtle spin splitting of the edges of the SOC-induced hybridisation gap. 
This splitting, seen in Fig.~\ref{fig:qpiwsoc}(b), results in shoulder-like features in simulated spectra that are similar to the experimental differential conductance spectrum\cite{marques_magneticfield_2021}, however, in the QPI experiments, this splitting can be linked to nematicity, while the loop current phase does not exhibit nematicity in the charge channel (see blue and red arrows in Fig.~\ref{fig:qpiwsoc}(d)).
We show that even for staggered moments $m$, this spin-splitting results in a spin-polarisation with magnitude large enough to be detected by SP-STM, on the order of $20\%$. The resulting checkerboard order, Fig.~\ref{fig:fig4}(d), has the same periodicity as the checkerboard order discussed in the literature\cite{matzdorf_ferromagnetism_2000, marques_magneticfield_2021}, however, the experimentally observed order occurs in the charge channel rather than in the spin polarization as in the model here (Fig.~\ref{fig:fig4}), and it is currently not known whether it is spin-polarized with indications from $\mu$SR for some magnetism in the near surface region\cite{fittipaldi_unveiling_2021}. Future SP-STM measurements can be directly compared to our calculations and will allow to verify the presence of loop currents and whether they are at the basis of the checkerboard order.

The loop current order considered here breaks a mirror symmetry of the unit cell, resulting in spin-splitting of the edges of the SOC gap. Other loop current orders proposed in the literature can result in breaking of inversion symmetry, which would result in Rashba spin-splitting. This Rashba-type spin-splitting would be staggered within the unit cell and likely results in new emergent phenomena.

Our results establish the expected experimental signatures of loop current orders in spectroscopic measurements. For the strength of loop current magnetic moments reported in the literature, clear signatures are expected in tunneling spectroscopy. We identify how loop currents affect the electronic structure of models without and with spin-orbit coupling, providing guidance on what signatures of loop currents to expect with a conventional tip in the first case, and with a spin-polarised tip in the latter case. 

The formalism can easily be generalised to other materials. While here, we only consider one specific realization of a loop current order, we anticipate that for other types of orders (for example intra-unit-cell orders), the order of magnitude of the effects of loop currents would remain the same, however opening the resulting hybridization gaps in other regions of the Brillouin zone depending on the ordering moment $\textbf{q}$, with a different prefactor due to a different enclosed area, and potentially inducing Rashba-spin-splitting for orders that break inversion symmetry. Our results provide guidance for future experiments seeking to verify evidence for loop current orders from STM measurements.

\begin{acknowledgments}
CAM, LCR and PW acknowledge funding from the Leverhulme Trust through Research Project Grant RPG-2022-315, and  CAM and PW from UKRI1107. This work has used computational resources of the high-performance computing cluster
Hypatia at the University of St Andrews.
VM, LCR performed initial model development, with calculations performed by VM, PW and CAM. CAM supervised the project. VM, PW and CAM prepared the figures and wrote the manuscript with contributions from all authors. All authors discussed the manuscript.
\end{acknowledgments}

\label{Bibliography}
\bibliographystyle{unsrtnat}
\bibliography{loopcurrents}

\setcounter{section}{0}
\renewcommand*{\theHsection}{S\the\value{section}}
\renewcommand\thesection{S\arabic{section}}
\makeatletter
\renewcommand{\c@secnumdepth}{0}
\makeatother
\section{Supplementary: Methods}\label{supp:Methods}

A realistic calculation of tunneling spectra requires accounting for the wave function overlap between the electronic states in the sample and the tip. To achieve this, we employ the continuum Green's function method\cite{choubey2014,Kreisel2015,kreisel_towards_2016,kreisel_quasi-particle_2021} as implemented in CalcQPI\cite{wahl_calcqpi_2025,wahl_calcqpi_code_2025} using tight-binding models and localized wave functions informed by Density Functional Theory calculations to compute the continuum local density of states (cLDOS) 

\begin{equation}
    \rho(\mathbf{r},\varepsilon)=-\mathrm{Im}G(\mathbf{r},\mathbf{r},\varepsilon)
\end{equation}

as a function of real-space coordinates $\mathbf{r}$ and energy $\varepsilon$ at a certain height $z$ of the tip above the top-most atom in the surface. We assume a point-like impurity potential $V_{\mathrm{imp}}$, acting equally on each orbital.

For the one-band cuprate model, we used the Wannier-projected tight-binding model of \ce{Bi2Sr2CaCu2O8} and the Cu $d_{x^2-y^2}$ Wannier function from A. Kreisel \textit{et al.}~\cite{Kreisel2015}, including renormalization and chemical potential shift. The cLDOS calculations were performed with an oversampling of 4 points per unit cell and window of 3 unit cells, at a $z$-height of \qty{1.6}~\AA~above the topmost atom of the unit cell, corresponding to the \ce{BiO} surface. The cLDOS was calculated over $84\times84$ unit cells in real space, and using $840\times 840$ $k$-points, with an energy broadening of \qty{1}{\milli\eV}. A single point-like impurity was introduced in the unit cell at the center of the real-space image, with an impurity potential of $V_{\mathrm{imp}}=\qty{0.3}{\eV}$. 

For the calculations for \ce{Sr2RuO4}, we used a Wannier-projected tight-binding model from a DFT calculation of a free-standing layer of \ce{Sr2RuO4}, including rotations of the \ce{RuO6} octahedra with an angle of $9^\circ$, corresponding to a model of the surface layer of a \ce{Sr2RuO4} crystal, as previously employed in ref.~\onlinecite{chandrasekaran_engineering_2024,profe_magic_2024}. The Wannier functions are the projected functions from Wannier90 for the same calculation, taken at a constant height of \qty{3}~\AA~above the \ce{SrO} surface. 
We add spin-orbit coupling to the tight-binding model to the $t_{2g}$ manifold, with constant $\lambda=\qty{0.15}{\eV}$.
The cLDOS calculations were performed with $1024\times1024$ $k$-points. For the QPI calculations in Fig.~\ref{fig:fig2}, \ref{fig:qpiwsoc}, a broadening of \qty{0.2}{\milli\eV} has been used, with a lattice of $64\times 64$ unit cells, oversampling of $4$ and a window of $2$. The scattering potential has been set to $V_0=\qty{0.5}{\eV}$. For the panels in Fig.~\ref{fig:fig4}, the calculations have been done over $8\times8$ unit cells in real space, with oversampling of $32$ and a window of $2$. The energy broadening was \qty{1}{\milli\eV}.

For all models, the staggered loop current order was introduced in the model via Eq.~\ref{eq:hamilton}, keeping the bandwidth constant by keeping $|t|$ constant as described in the text.

\end{document}